\documentclass[12pt]{iopart}

\usepackage{iopams}
\usepackage{graphicx}
\begin{document}

\title[]{Interference of electrons in backscattering through a quantum point contact}

\author{A.~A.~Kozikov, C.~R\"{o}ssler, T.~Ihn, K.~Ensslin, C.~Reichl and W.~Wegscheider}

\address{Solid State Physics Laboratory, ETH Z\"{u}rich, CH-8093 Z\"{u}rich, Switzerland}
\ead{akozikov@phys.ethz.ch}
\begin{abstract}
Scanning gate microscopy is used to locally investigate electron transport in a high-mobility two-dimensional electron gas formed in a GaAs/AlGaAs heterostructure. Using quantum point contacts (QPC) we observe branches caused by electron backscattering decorated with interference fringes similar to previous observations by Topinka et al. \cite{TopinkaSci}. We investigate the branches at different points of a conductance plateau as well as between plateaus, and demonstrate that the most dramatic changes in branch pattern occur at the low-energy side of the conductance plateaus. The branches disappear at magnetic fields as low as 50 mT demonstrating the importance of backscattering for the observation of the branching effect.
%and the interference fringes is shown in
The spacing between the interference fringes varies by more than 50\% for different branches across scales of microns. Several scenarios are discussed to explain this observation.
\end{abstract}

%\pacs{00.00, 20.00, 42.10}

\maketitle

\section{Introduction}

Scanning gate microscopy (SGM) is a scanning probe technique, in which the conducting tip of an atomic force microscope acts as a moveable gate locally changing the electron density beneath it. It allows investigating the electron behavior beyond conventional transport measurements. Its successful implementation started about 16 years ago \cite{Eriksson} and was followed a few years later by the pioneering experiment to image the electron backscattering through a narrow constriction in a high-mobility two-dimensional electron gas (2DEG) \cite{TopinkaSci,TopinkaNat}. Since then the SGM technique has become widely used for local studies of electron transport in different nanostructures \cite{Woodside,Pioda2004,Fallahi,Aoki,Hackens,Pioda,Gildemeister,Bleszynski,SchnezPRB1}.

Back in the year 2001 when electron backscattering through the QPC was imaged by Topinka and coworkers \cite{TopinkaNat}, it was unexpectedly found to be dominant along narrow branches decorated by interference fringes \cite{TopinkaNat} about half the Fermi wavelength apart. The branching behaviour has been studied for several years since then \cite{LeRoyDensity,JuraNat,JuraPRB1,Paradiso,JuraPRB2}. It was found that classical mechanics could explain the formation of branches \cite{TopinkaNat}, whereas quantum mechanics was needed to account for their stability upon initial conditions \cite{JuraNat}. Local electron transport through a QPC was used to study electron-electron interactions \cite{JuraPRB2} or to map the local carrier density \cite{LeRoyDensity}.

The branching effect occurs due to focusing of electron waves by small-angle scattering off a random background potential created by ionized doping atoms in the heterostructure. The interference fringes close to the QPC appear because of the interference of electron waves coming from the QPC to the tip-depleted region and those scattered back to the constriction. A few microns away from the QPC such interference may not survive either due to electron dephasing or due to thermal averaging, or due to both. In this case interference of electron waves scattered between the tip and closely located sharp impurity potentials becomes essential.

Imaging electron backscattering is possible due to the effect the tip has on the 2DEG. By applying a sufficiently negative voltage to the tip, a depleted region in the 2DEG appears, which acts as a backscatterer. Electron waves leaving the QPC are scattered back to the constriction by the potential induced by the tip. When the tip is placed in a way that it interrupts parts of the flow from a particular QPC mode, the transmission of this mode decreases. This is seen as a change in the conductance across the sample, which is measured as a function of the tip position allowing imaging electron backscattering through the QPC.

In this paper we show in detail how the QPC conductance is affected by the tip position, tip bias and low magnetic fields. We study details of the local potential landscape, which can be inferred from the spacing between interference fringes.
%In this paper we continue deepening our knowledge of electron flow through the QPC and studying details of the local potential landscape, which can be inferred from the spacing between interference fringes.
Our sample has a mobility which is about a factor of two higher than the best previously studied samples \cite{JuraPRB1}. We find that the variations of the spacing of the interference fringes can reach 50$\%$ on a scale of a few hundred nanometers independent of the tip used, top gate voltage and thermal cycles. We discuss several scenarios to explain these variations and it turns out that each consideration has its shortcomings. These scenarios are based on a locally varying or constant carrier density. We investigate how the behaviour of  backscattering changes as a function of the top gate voltage by taking several images on a conductance plateaus as well as between the plateaus. Studies of the branches of electron backscattering and interference fringes in a perpendicular magnetic field highlight that backscattering is indeed required for the observation of the branched flow. In order to show the functionality and quality of our device we show a series of experiments in the appendix (\fref{fig:Asym}) which have been performed in a similar way by Topinka and collaborators in a number of papers.

\begin{figure}[t]
\begin{center}
\includegraphics[width=\columnwidth]{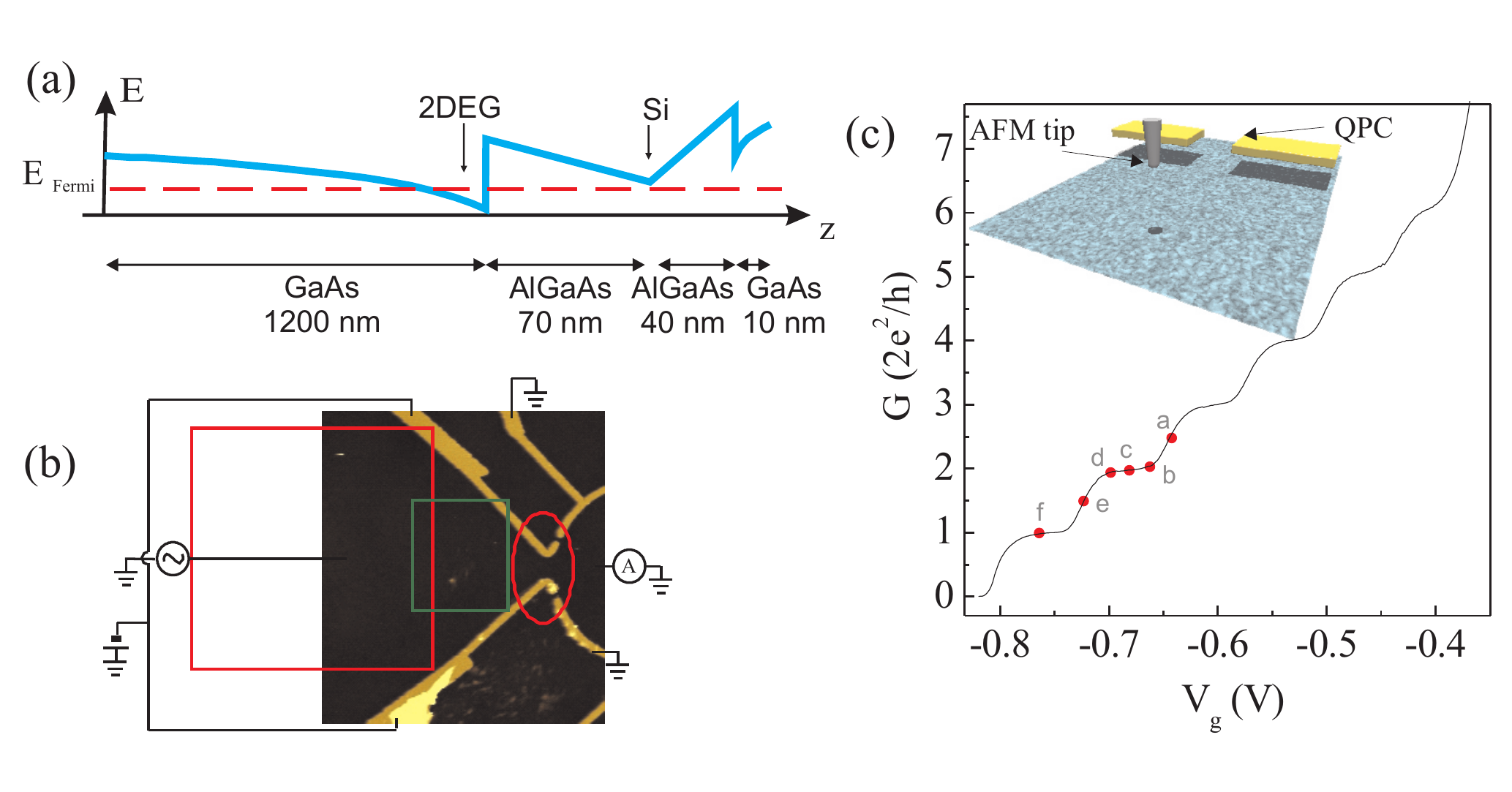}
\caption{(a) Schematics of the heterostructure layers. The z-axis shows the direction of growth. The blue line corresponds to a profile of the conduction band edge and the dashed line to the Fermi energy. The 2DEG is located at the AlGaAs/GaAs interface 120 nm below the surface. Silicon (Si) donor atoms ($3.2\times10^{12}$~cm$^{-2}$) are separated from the 2DEG by the AlGaAs spacer 70 nm thick. (b) An image of the investigated sample obtained at room temperature by a commercial AFM. The dark area is the surface of the GaAs heterostructure and yellow contacts are top gates. The QPC used to study electron backscattering is marked by a red circle. All other top gates are grounded. Conductance across the sample as a function of the tip position is imaged at the location of the red and green squares. The right-hand side borders of the squares are about 1.5 and 0.5 $\mu$m away from the center of the constriction, respectively. The sizes of the squares are 3.5$\times$3.5 and 1.7$\times$2.0 $\mu$m$^2$. (c) Gate voltage dependence of the conductance measured in a two-terminal configuration at about 300 mK. Plateaus spaced by $2e^2/h$ are clearly seen. The contact resistance has been subtracted. Red circles indicate the gate voltages at which electron backscattering is imaged. Letters a-f by these circles correspond to the graphs in \fref{fig:Gate}. The insert shows a schematics of the AFM tip scanning over the surface of the GaAs heterostructure. When negatively biased the top gates and the tip create depleted regions (shaded ares) in the 2DEG with the random background potential.}
\label{fig:one}
\end{center}
\end{figure}

\section{Experimental methods}

The two-dimensional electron gas (2DEG) under investigation is embedded in a GaAs/AlGaAs triangular quantum well structure with an aluminum content of 24.4$\%$ (\fref{fig:one}(a)). It is located 120 nm below the surface and has a mobility of around $8\times10^6$ cm$^2$/Vs and an electron density $n = 1.2\times10^{11}$ cm$^{-2}$ determined from Hall effect measurements at 300 mK. The corresponding Fermi wavelength and the elastic mean free path are $\lambda_\mathrm{F}=72$ nm and $l_\mathrm{p}=49~\mu$m, respectively, and the Fermi energy is $E_\mathrm{F}=4.4~$meV. Silicon donor atoms are located in a plane 70 nm above the 2DEG. A list of all quantities relevant for the experiment is given in Table \ref{tab:Parameters}.

\begin{table}[b]
\caption{Parameters of the sample at 300mK. Here $l_{\mathrm{th}}=(hD/k_B T)^{0.5}$ and $l_\varphi$ are the thermal and dephasing length, respectively, and $V_{\mathrm{tip}}$ is the tip bias.}
\begin{tabular}{p{1.3cm}p{2.0cm}p{2.0cm}p{2.2cm}p{2.1cm}p{2.3cm}p{0.8cm}}
%\begin{tabular}{c|ccc|c|c}
\hline\hline
%\multicolumn{1}{c}{} & \multicolumn{3}{c}{S1} & \multicolumn{1}{c}{S2} & \multicolumn{1}{c}{S3}\\
\hline
$\lambda_\mathrm{F}$ (nm)\centering&$E_\mathrm{F}$\centering\par (meV)&$\mu$\centering \par(m$^2$/Vs)& $n$ ($10^{11}\centering$~cm$^{-2}$)\centering&$l_\mathrm{p}$\centering \par($\mu$m)&$l_{\mathrm{th}}$\centering \par ($\mu$m)& ~$V_{\mathrm{tip}}$\ \par ~(V)\\
\hline
72\centering &   4.4\centering  & 800\centering  &  1.2\centering & 49\centering &9\centering&~-4.5\\
\hline\hline
$l_\varphi$ \centering\par($\mu$m)&QPC width (nm)\centering&2DEG depth\centering \par(  nm)&Gate height (nm)\centering&Spacer thickness (nm) \centering&Tip-surface distance\centering \par(nm)&\\
\hline
200\centering & 300\centering &120\centering &30\centering&70\centering&60\centering&\\
\end{tabular}
\label{tab:Parameters}
\end{table}

All structures are defined electrostatically using Ti/Au gates (height 30 nm) on top of the GaAs surface (\fref{fig:one}(b)). Due to the presence of the gaps between the gates, it is possible to use only some of the structures shown in the figure. Experiments are performed with the quantum point contact (QPC) encircled in \fref{fig:one}(b). The lithographic width of the constriction is around 300 nm. All other top gates are grounded.

\begin{figure}[t]
\begin{center}
\includegraphics[width=15cm]{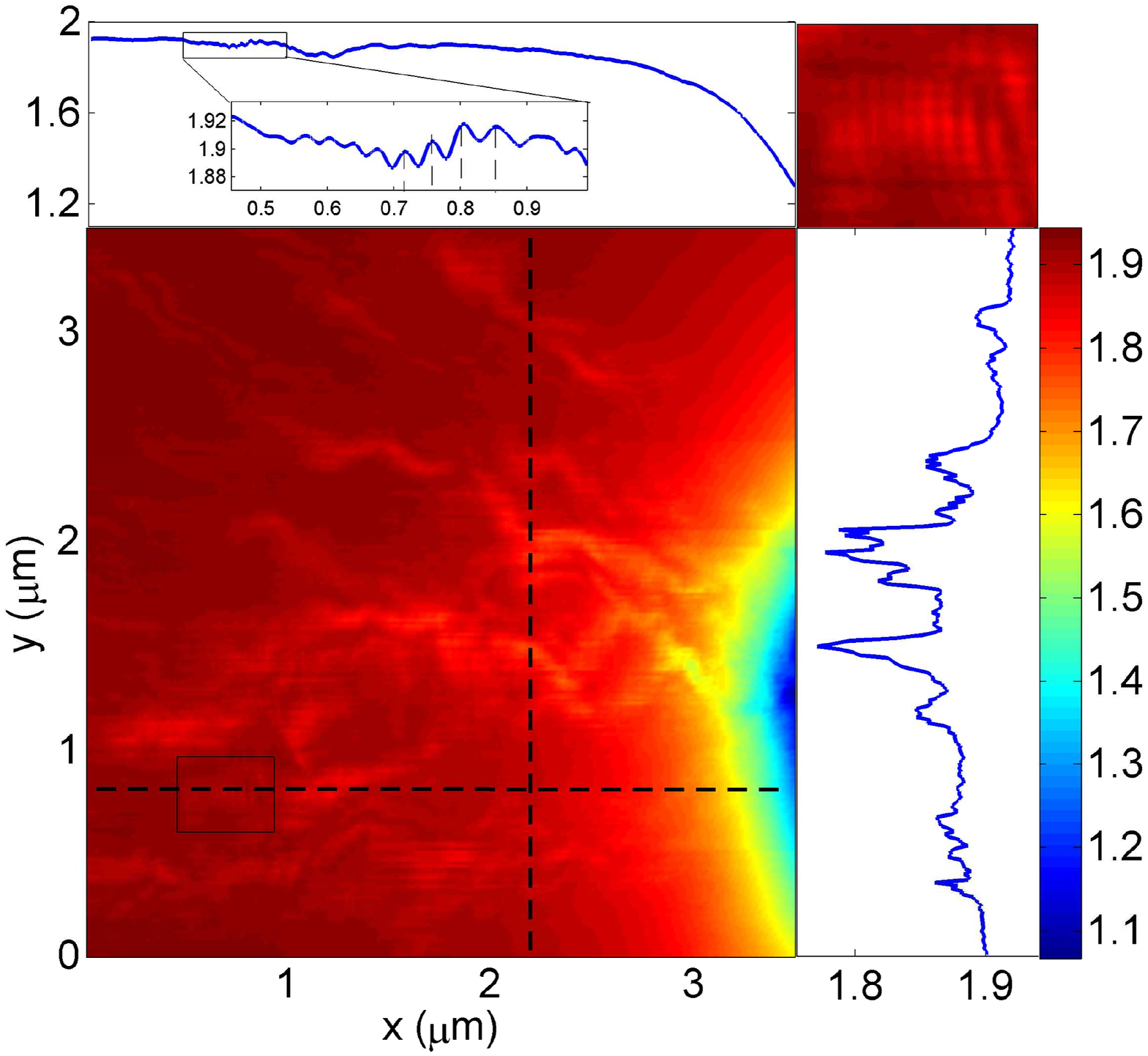}
\caption{The QPC conductance as a function of the tip position in units of 2$e^2/h$ (see the colour scale) measured in the area indicated by the red square in \fref{fig:one}(b). The 1D plot at the top (on the right-hand side) is a cross-section along the horizontal (vertical) dashed line. A small colour plot in the upper right corner is a zoomed-in region shown in the main graph by a rectangle. The colour scale is slightly different compared to the main graph to make the interference fringes more visible. These fringes are also seen in a zoomed-in region in the horizontal cross-section.}
\label{fig:RawData}
\end{center}
\end{figure}

\begin{figure}[t]
\begin{center}
\includegraphics[width=15cm]{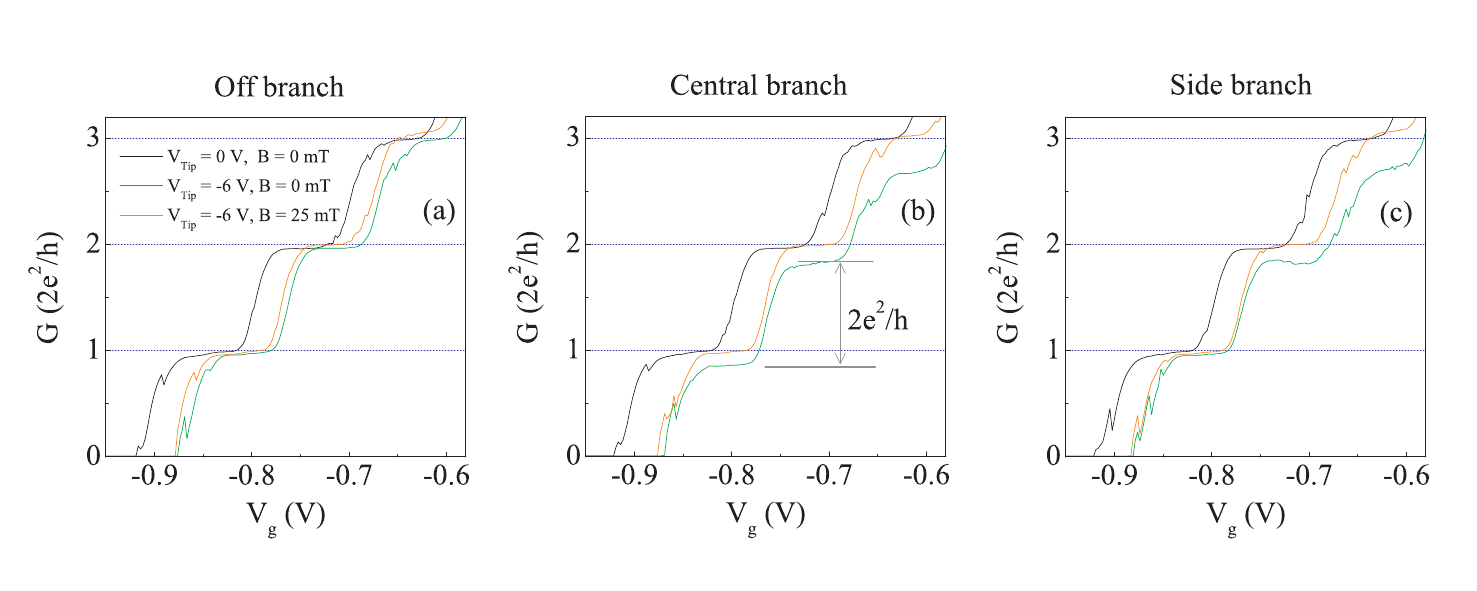}
\caption{ The $G(V_\mathrm{g})$ curves measured when the tip is placed away from branches (a), above a central branch (b) and above a side branch (c) at different tip biases, 0 and -6~V, and magnetic fields, 0 and 25~mT.}
%The effect of the magnetic field on the QPC conductance is shown in (b), (d) and (f). Traces at $V_\mathrm{tip}=0$ and $-6$ V at $B=0$ mT are the same as in the upper panel. The two other curves correspond to $B=15$ and 25 mT.}
\label{fig:GVgTip}
\end{center}
\end{figure}

The differential conductance across the sample, $G~=~\rmd I/\rmd V_{\mathrm{SD}}$, is measured using a standard lock-in technique in a two-terminal configuration by applying an AC rms source-drain voltage of 100 $\mu$V.
The QPC is formed (\fref{fig:one}(c)) when the 2DEG beneath the gates is depleted at $V_\mathrm{g}=-380~$mV. (As seen from \fref{fig:one}(b) the width of the top gates varies. The electron gas is first depleted beneath the wide part at around -250~mV, which is close to the value estimated using a parallel plate capacitor model, and then beneath the narrow part at -380~mV, for which this capacitor model is not valid). At more negative gate voltages the width of the constriction decreases as well as the conductance through it. Six plateaus spaced by 2$e^2/h$ are clearly seen (\fref{fig:one}(c)) until the QPC pinches off at $V_\mathrm{g}=-830~$mV.

SGM experiments are performed at a base temperature of 300 mK using a home-built atomic force microscope in a He-3 system \cite{Ihn}. The differential conductance, $G$, across the sample is measured as a function of the tip position $(x,y)$. Unless stated otherwise, the Pt/Ir tip is placed 60 nm above the GaAs surface and biased to -4.5 V. Most of the images presented here are obtained by scanning an area of $3.5\times3.5$ $\mu$m$^2$ located about 1.5 $\mu$m to the left of the QPC (red square in \fref{fig:one}(a)). Red circles in \fref{fig:one}(c) indicate gate voltages at which the electron backscattering is imaged.

\Fref{fig:RawData} shows the QPC conductance (the same contact resistance as in \fref{fig:one}(c) has been subtracted) as a function of the tip position in the middle of the second conductance plateau (see \fref{fig:one}(c)). One can see that in the largest part of the image the conductance is very close to $4e^2/h$ meaning that the tip has a very small effect on the QPC conductance. In this region there are narrow branch-like areas (bright red) where the conductance decreases. A cross-section along the vertical dashed line (the blue curve on the right-hand side) shows a series of dips in $G$ that correspond to the bright spots along the dashed line. When the tip moves closer to the QPC, the conductance drops to about $1.2\times e^2/h$. A 1D cross-section at the top indicates how $G$ varies along the horizontal dashed line: $G$ remains almost unchanged until the tip comes closer to the QPC causing a rapid decrease in $G$. One can also see wiggles on this curve (a zoomed-in region shown as an inset). These wiggles are also seen in the 2D colour plot (upper right corner), which is a zoomed-in region of the main graph indicated by a rectangle.

Thus, during scanning there are changes of the conductance occurring on distinctly different scales: 1) an overall change in $G$ on a length scale of microns (the curve at the top) and on a conductance scale of $2e^2/h$, 2) there are branch-like areas on a length scale of 100 nm (the curve on the right-hand side) and on a conductance scale of $0.1\times2e^2/h$ and 3) the wiggles on a length scale of the Fermi wavelength (the curve at the top) and on a conductance scale of $0.01\times2e^2/h$. Following the existing notation \cite{TopinkaSci, TopinkaNat} we call the wiggles interference fringes. Below we discuss possible origins of their formation based on our experimental results. These are the effects that the negatively biased tip has on the QPC conductance.

\begin{figure}[t]
\begin{center}
\includegraphics[width=15cm]{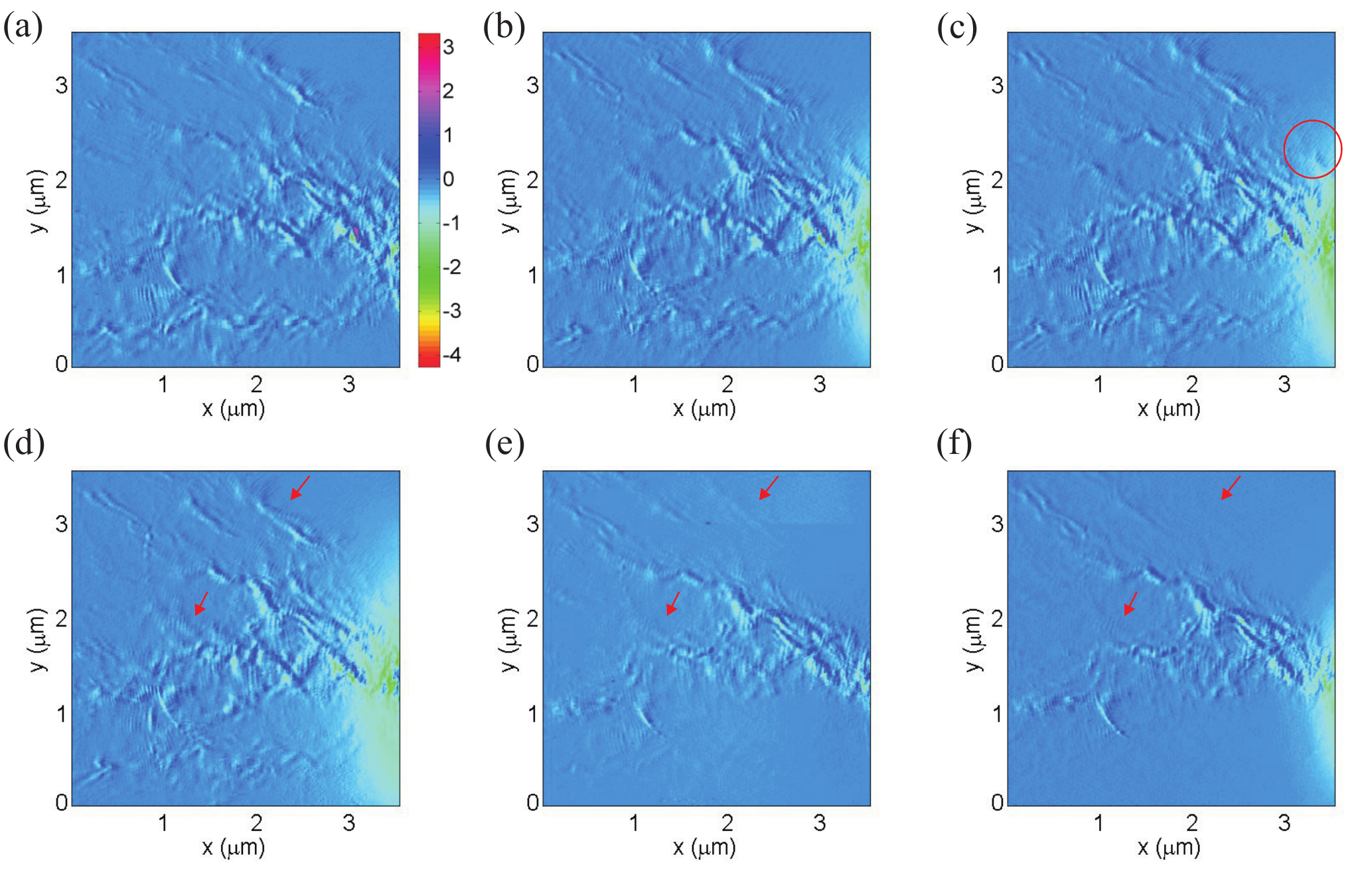}
\caption{(a)-(f) Numerical derivative of the conductance, $\rmd G/\rmd x$, plotted in colour scale as a function of tip position at different gate voltages shown by red circles in \fref{fig:one}(c): (a) $V_\textrm{g}=-0.655$V, (b) $V_\textrm{g}=-0.675$V, (c) $V_\textrm{g}=-0.684$V, (d) $V_\textrm{g}=-0.690$V, (e) $V_\textrm{g}=-0.725$V, (f) $V_\textrm{g}=-0.750$V. Red arrows in (f)$\rightarrow$(d) indicate the appearance of new branches during a transition from the first conductance plateau to the second. The colour scale has units of $10^{-6}$ S/m and it is the same for each plot.}
\label{fig:Gate}
\end{center}
\end{figure}

To exhibit the effect of the tip-induced potential on the QPC conductance more clearly we present a set of the $G(V_\mathrm{g})$ curves measured as a function of the tip position, tip bias and magnetic field (\fref{fig:GVgTip}). When the tip is placed above the branch-free areas, \fref{fig:GVgTip}(a), and the tip bias is gradually changed from 0 to -6 V, the conductance at the plateaus does not change, but the entire curve shifts towards positive gate voltages. Therefore, when the tip scans the surface (even microns away from the QPC) at a fixed $V_\mathrm{g}$, the conductance decreases. The tip acts as a top gate to regions remote from the directly depleted area below the tip. Its induced potential changes the local potential of the constriction and thereby decreases the conductance across the sample. We call the horizontal shift of the $G(V_\mathrm{g})$ curve the gating effect. This is consistent with the horizontal cross-section in \fref{fig:RawData}: at the largest distance from the QPC the biased tip reduces the conductance to about $1.9\times2e^2/h$. As it moves closer to the QPC, $G$ does not change much, because the gradient of the tip-induced potential at the QPC is small (tails of a Lorentzian). Close to the constriction this gradient is large, therefore the conductance decreases rapidly. Applying a perpendicular magnetic field (\Fref{fig:GVgTip}(a)) does not change the $G(V_\mathrm{g})$ curves as expected.

\Fref{fig:GVgTip}(b) shows the situation when the tip is placed above the central branch (in front of the QPC). Again we see that the curve shifts towards positive gate voltages (gating effect) when the tip bias reaches -6~V. But apart from that, the plateau conductance decreases, because the tip depleted region scatters electrons back to the constriction reducing the transmission of a particular QPC mode leading to the decrease of the conductance. The central branch is populated by electrons from QPC modes with odd quantum numbers in confinement direction. That is why the step heights from pinch-off to the first and from the second to the third plateaus are smaller than $2e^2/h$ and that from the first to the second plateau equals $2e^2/h$. As one can see, $G$ is most sensitive to electron backscattering when the tip is placed above a branch. This branching effect was shown in the vertical cross-section in \fref{fig:RawData}. The magnetic field suppresses backscattering and, therefore, increases the conductance, \fref{fig:GVgTip}(b). At 25 mT the conductance of all plateaus is restored. This can also happen if the Lorentz force bends electron trajectories in a way that the electrons do not reach the tip.

When the tip is placed above a side branch, \fref{fig:GVgTip}(c), all plateaus apart from the first one are affected. This is due to the angular distribution of electrons leaving the constriction \cite{TopinkaSci}, which depends on the number of open QPC modes. As seen from the height of the plateaus in the figure, the second and the third modes fill the side branch. A magnetic field of 25 mT is enough to restore the conductance.
The magnitude of the magnetic field sufficient to restore the plateau conductance depends on how far away the tip is from the QPC in comparison with the cyclotron radius.

As seen from \fref{fig:RawData} the interference fringes are almost not seen in raw data. Therefore, we plot the differentiated conductance, $\rmd G(x,y)/\rmd x$, in further images. This way the strong effect of the background will be eliminated and the interference fringes are enhanced.

\Fref{fig:Gate}(a)-(f) shows the evolution of the branches at several closely spaced values (\fref{fig:one}(c)) of the gate voltage. One can see that when the value of the conductance through the QPC is not an integer and lies above a plateau (\fref{fig:Gate}(a)), as well as anywhere on the plateau, \fref{fig:Gate}(b)-(d), the branching pattern is almost the same. When the conductance is just below a plateau value, the pattern changes significantly (\fref{fig:Gate}(d) to (e)). Such a trend is observed for other plateaus, e.g. for the first plateau (\fref{fig:Gate}(e) and (f)). In other words, when the source and drain electrochemical potentials lie between two neighboring minima of the subband energy dispersion relations, the observed branch pattern stays the same. As more and more modes contribute to the conductance as we increase the QPC gate voltage, the pattern changes and branches can be observed in a larger angular range. A transition from the middle of the first plateau to the onset of the second is accompanied with appearing additional branches shown by red arrows in (f)$\rightarrow$(d).
Each branch in (a)-(f) is decorated with interference fringes mostly running perpendicular to the direction of electron backscattering. (A larger size of the plots is required to clearly see them. An example is given in the supplementary material). The origin of the fringes will be discussed in detail below. In previous experiments several models were proposed to explain the fringe formation \cite{JuraNat, JuraPRB1}. Here we have to take into account the higher mobility of our sample and the tip geometry and size.

The experimental conditions, such as the tip bias and the tip-surface distance, used in our experiments, were experimentally found to be sufficient to create a depleted region in the 2DEG and, therefore, to observe interference fringes and the branching effect. The tip placed 60 nm above the surface starts depleting the electron gas when biased at about -2.7 V. At -4.5 V the size of the depleted region is of the order of 1 $\mu$m.  A comparison between different tip-surface distances and tip biases is given in the supplementary material. In previous studies of electron backscattering, SGM images were obtained by placing the tip about 10-30 nm above the surface \cite{JuraPRB1}. The lateral tip-QPC distance in our experiments, $L$,  fulfils the relations $\lambda_\mathrm{F}\ll L \ll l_\mathrm{p}$ and $L\lesssim l_{\mathrm{th}}$. Scanning more than a micron away from the constriction is important to be less affected by the gating effect.
 %and scanning at distances from the QPC, which were larger than the thermal length, $l_{\mathrm{th}}$ (about three times shorter than in our experiment \cite{JuraPRB1}). The observed interference fringes resulted from interference of waves, whose path lengths were either equal to each other or smaller than $l_{\mathrm{th}}$. In our case the thermal length is larger than the distance between the tip and the QPC. This may result in interference of multiply reflected electron waves.

\begin{figure}[t]
\begin{center}
\includegraphics[width=15cm]{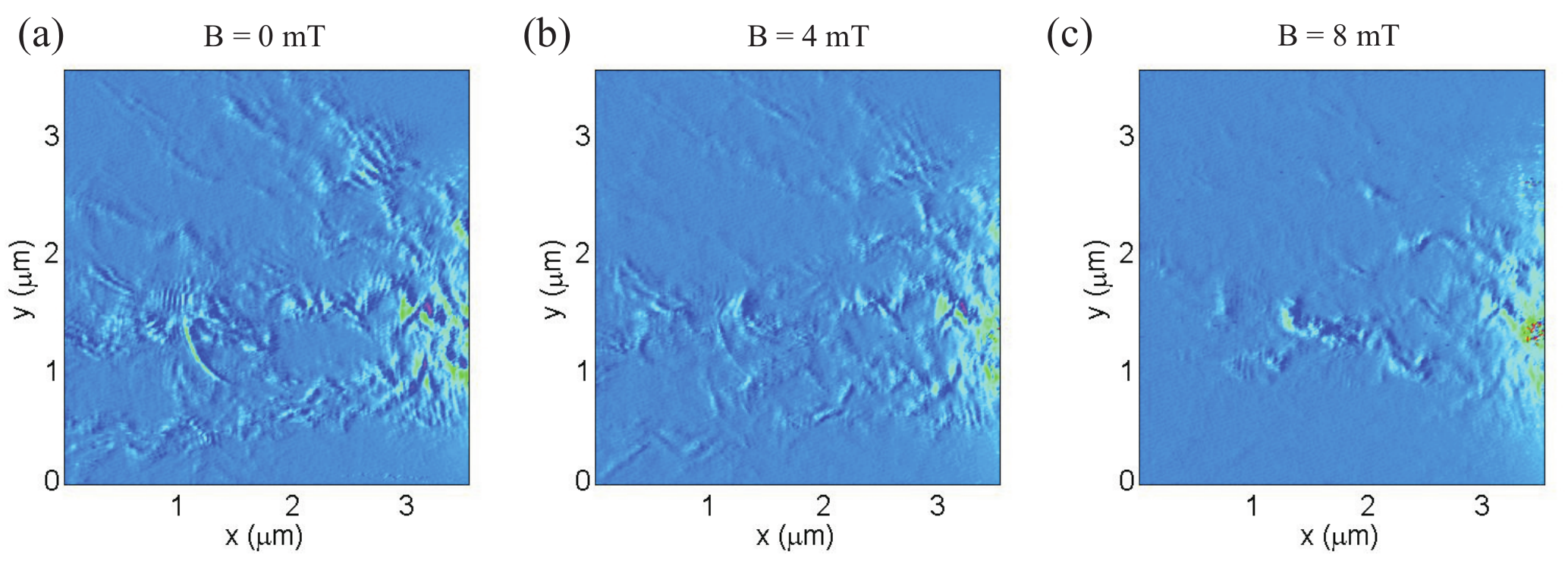}
\caption{Numerical derivative of the conductance, dG/dx, plotted in colour scale as a function of tip position for different magnetic fields. The conductance across the sample is $G=6e^2/h$. (a) B = 0 mT, (b) B = 4 mT, (c) B = 8 mT. The values for the cyclotron radius in (b) and (c) are 14 and 7 $\mu$m, respectively.}
\label{fig:Bfield}
\end{center}
\end{figure}

One can also see from \fref{fig:Gate} that it is difficult to distinguish between different branches in the area closest to the QPC. This happens due to the gating effect that becomes stronger when the tip moves closer to the constriction. Later we show how images change when this effect is compensated.

Previously we showed how the magnetic field affects the QPC conductance as a function of the gate voltage at different positions of the tip. In \fref{fig:Bfield} we image this effect when the conductance is set to $3\times2e^2/h$. One can see that as the magnetic field increases from 0 to 8 mT the branches and the fringes gradually disappear. Although they are still seen at 8 mT, applying a higher magnetic field makes both of them fade away completely. These images together with the $G(V_\mathrm{g})$ curves in \fref{fig:GVgTip} illustrate again that backscattering is an essential part of the involved phenomena. As seen from the images some branches and fringes disappear faster than others. The effect of the magnetic field on the QPC conductance at different positions of the tip is shown in \fref{fig:GVgTip}. This indicates that the mechanism of their formation is different in different parts of the scan area. For example, in case of multiple scattering off the tip and impurities, interference can survive due to the presence of different possible paths that electron waves can move along without losing their coherence (because the thermal length is larger or on the order of the tip-QPC distance). The upper and lower branches in \fref{fig:Bfield}(c) disappear already at 8 mT. A cyclotron radius of about 7 $\mu$m corresponds to this magnetic field. Therefore, these branches and fringes are likely caused by direct backscattering since they are located 2-4 $\mu$m away from the QPC, which is comparable to the cyclotron radius. In the center of the scan area the main mechanism of branch and fringe formation seems to be multiple scattering off the tip, impurities and top gates. One can also see in \fref{fig:Bfield}(b)-(c) that at some positions of the tip the change in the conductance becomes stronger, which is in contract to the fact that the magnetic field restores the conductance at the plateaus and therefore its change should eventually become zero. This may happen when electron trajectories bent by the Lorenz force come back to the QPC after scattering off the tip. These events, however, can be quite random due to the presence of the background potential.

\begin{figure}[t]
\begin{center}
\includegraphics[width=15cm]{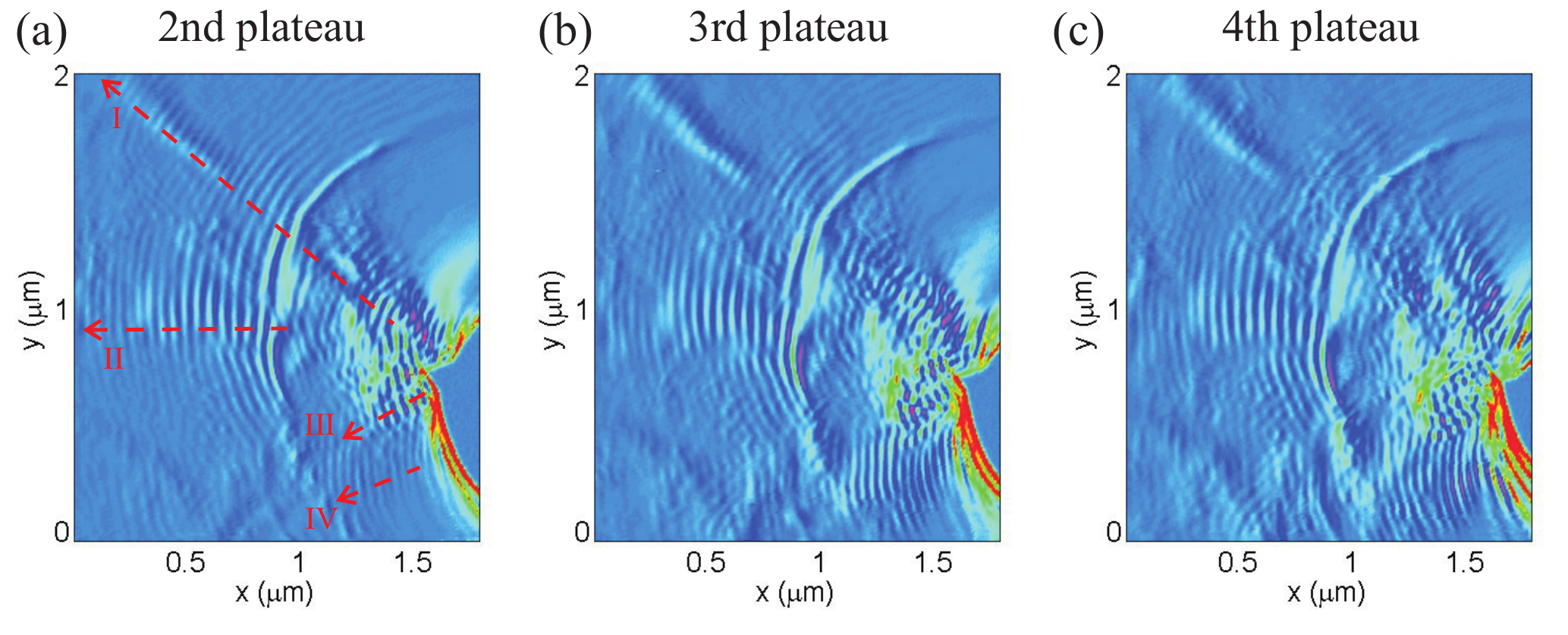}
\caption{Electron backscattering through the QPC imaged using a gating compensating technique when the QPC is biased on the second (a), third (b) and the fourth (c) plateau. Dashed arrows indicate lines and directions along which 1D profiles are analyzed (shown only for the second plateau). }
\label{fig:GateCompen}
\end{center}
\end{figure}

In the images shown so far the gating effect has been significant. It prevents observing details of the electron backscattering close to the point contact. In order to scan near the QPC and to increase the resolution, we partially compensate the gating effect in the following by performing a two-pass technique. In the first pass the tip is lifted by 120 nm above the surface and scanned at a constant height. At the same time the conductance through the QPC is kept constant by varying and recording the voltage applied to the gates, which is done by an additional feedback loop. At 120 nm the tip biased at -4.5 V does not deplete the electron gas beneath it (see Supplementary Material). Therefore, only the gating effect is present. In the second pass the tip is lowered to 60 nm above the surface and the conductance through the QPC is measured. The top gate voltage is changed according to the recorded data from the first pass. The result is shown in \fref{fig:GateCompen}. The scan area is slightly reduced and shifted by about 1 $\mu$m closer to the QPC (green box in \fref{fig:one}(a)).

The gating effect depends on the tip-surface distance close to the constriction, which can be seen in the figure. The current through the QPC drops to zero in the region closest to the constriction (an almost rounded region of the same colour). However, away from it the gating effect is fully compensated leading to a clear observation of interference fringes. The region of zero current can also be due to blocking any electron flow by the large tip before electrons reach the QPC. There is a small overlap between images in this figure and those in \fref{fig:Gate}. The upper branch in \fref{fig:Gate}(c) (encircled) is a continuation of that in \fref{fig:GateCompen}(a) (see dashed line I). All other features in \fref{fig:GateCompen} become visible only after compensating the gating effect. The interference fringes appear more pronounced than in previous images. They start at the region of zero current and run predominantly perpendicular to the radial directions. One can also see an area where fringes cross each other. Away from it, they form circular arcs (ring pattern) of different spacing between maxima/minima. Scanning closer to the QPC allows us to observe the widening of the interference pattern as a function of the QPC transmission even though the lobes are not seen well.

Thus, the gating compensation technique allows one to distinguish small changes in the conductance due to the backscattering effect by eliminating larger changes due to the gating effect. This essentially means that this technique shifts the $G(V_\mathrm{g})$ curves in \fref{fig:GVgTip} horizontally back to their original position at $V_\mathrm{tip}=0$ V. When the tip moves above a branch, only the changes in the conductance at the plateaus are seen in the images. The technique becomes very useful when scanning close to the QPC. A similar procedure to compensate the gating effect was reported in Ref. \cite{JuraPRB1}. There the authors also scanned the surface twice at different heights of the tip in order to detect only the effect of electron backscattering.

\section{Analysis of interference fringes}

\begin{figure}[t]
\begin{center}
\includegraphics[width=15cm]{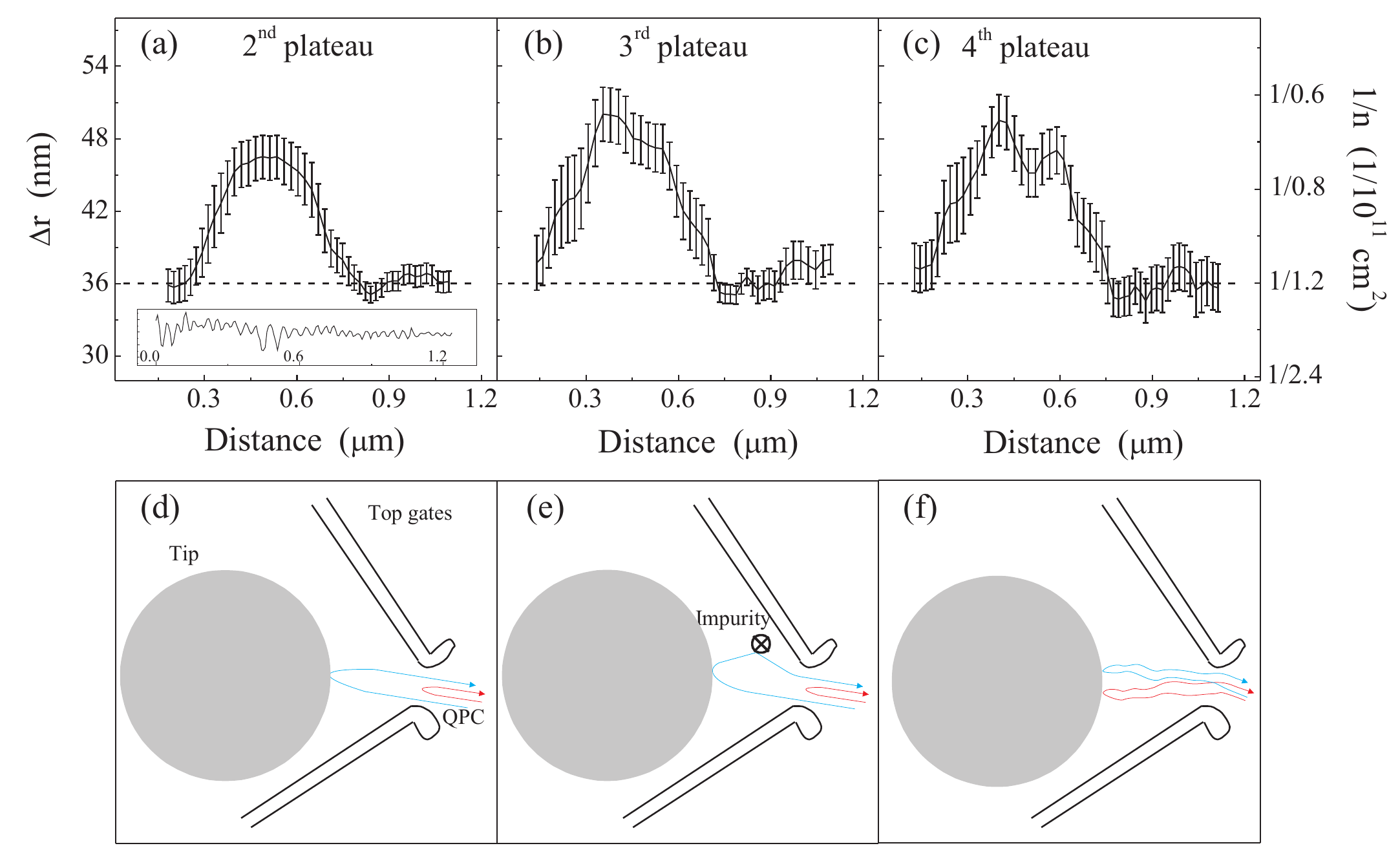}
\caption{(a)-(c) Left-hand side scale: the separation between interference fringes as a function of distance along the dashed line I (see \fref{fig:GateCompen}(a)) for the case of the second, third and the fourth conductance plateau. The dashed lines indicate half the Fermi wavelength. Error bars show the standard deviation of the mean. Right-hand side scale: the inverse variations of the extracted carrier density in units of $1/(10^{11}$ cm$^{-2})$ as a function of distance determined using equation \ref{eqn:LocalDensity} in scenario~1. The bulk carrier density is $1.2\times10^{11}$ cm$^{-2}$. (d)-(f) Explanations of the fringe spacing variations shown in (a)-(c) and the ring pattern: (d) scenario 1, (e) scenario 1 involving sharp impurities and (f) scenario 2, which takes into account the smooth but random background potential leading to the wavy electron trajectories.}
\label{fig:Spacing}
\end{center}
\end{figure}

The interference fringes, which decorate the branches, consist of a series of maxima and minima of $\rmd G(x,y)/\rmd x$.
The separation between maxima (minima) is determined from the profiles taken along the four dashed lines shown in \fref{fig:GateCompen}(a) for the three cases of the second, third and fourth conductance plateau. As an example, such a profile along line I is shown for the case of the second plateau in the inset of \fref{fig:Spacing}(a). As can be seen from this profile, the spacing between the fringes changes noticeably: the first half micron it increases and then decreases. The profiles are used to determine the separations between neighboring maxima (minima). For each such maximum and minimum of the profiles eight extrema from each side (17 points in total) are taken and the average over 17 data points gives one data point in the following graphs. For the dashed line I in figures \ref{fig:GateCompen}(b) and (c) (not shown) the averaging is performed over 11 points. For line III for all plateaus this is done over 7 points. The result of averaging is shown in \fref{fig:Spacing}(a)-(c) for the dashed line I. One can see that close to the constriction the spacing between the fringes is pretty accurately given by half the Fermi wavelength. Then it increases by more than 30$\%$ and decreases to the original value over a length scale of about 700 nm. This trend is independent of the transmission of the QPC. Although line I does not go through the phase gradient along its entire length, correcting its position in some regions will not change the qualitative and quantitative conclusion about strong deviation of the fringe spacing.

We consider several scenarios to explain the observed variations of the spacing between the fringes, $\Delta r$, and the ring pattern. In the first, we neglect the smooth background potential coming from the ionized donor atoms. Therefore, electron waves can be represented as geometrically straight lines. We also assume that the fringes appear due to interference of an electron wave directly reflected from the QPC and that transmitted through it and backscattered off the tip (\fref{fig:Spacing}(d)). In this case the phase difference between them is $\Delta\phi=2\int\limits_0^{\tilde{r}} \sqrt{2\pi n^*(r)}dr$, where $n^*(r)$ is the local carrier density at a distance $r$ from the QPC and $\tilde{r}$ is the distance from the constriction to the edge of the tip-depleted region. The phase difference becomes equal to an integer multiple of $2\pi$ between two neighboring maxima (minima), see the inset of \fref{fig:Spacing}(a). Thus, a spatially dependent spacing  $\Delta r$ of the interference fringes results in a local carrier density, $n^*(r)$:

\begin{eqnarray}
n^*(r)=\frac{\pi}{2\Delta r^2}
\label{eqn:LocalDensity}
\end{eqnarray}

This result does not depend on angle, because the interfering waves meet at 180$^{\circ}$ as was assumed. In this scenario it is also possible that the electron wave transmitted through the QPC scatters off the tip and a sharp impurity potential and returns back to the constriction (\fref{fig:Spacing}(e)). In this case the fringe spacing could be larger than $\lambda_\mathrm{F}/2$ since the interfering waves may include a finite angle. Experiments in samples with different levels of disorder confirm that in high-mobility samples the number of impurities with a short-range potential is very small \cite{JuraNat}. The fact that we see a strongly varying fringe spacing for a number of different experimental situations renders this particular case of scenario 1 rather unlikely.

In scenario 2 we assume that the local carrier density is constant in space and equal to the bulk density (\fref{fig:Spacing}(f)) everywhere. Interference occurs at the QPC between different partial waves backscattered from the tip, which meet at the constriction under some angle. This is possible due to the relatively large size of the tip-depleted region (on the order of a micron). The spacing between the fringes in this scenario depends on the angle and also on the details of the spatially modulated background potential between tip and constriction, without which the waves will not meet at the QPC after backscattering off the tip.

Variations of the carrier density extracted using equation \ref{eqn:LocalDensity} of scenario 1 along profile I are shown in \fref{fig:Spacing}(a)-(c) (right-hand side scale). One can see that the extracted carrier density at the beginning and at the end of the profile is the same as the bulk density determined from the Hall effect measurements ($n = 1.2\times10^{11}$ cm$^{-2}$). However, in-between it decreases to almost half the value. A similar analysis is done for profiles II-IV in \fref{fig:GateCompen}(a). The extracted density along line II is similar to that along line I: $n^*$ increases gradually from ~0.7 to $1.2\times10^{11}$ cm$^{-2}$. Along line III $n^*\thickapprox1.2\times10^{11}$ cm$^{-2}$ and along line IV it is 25$\%$ higher. Considering the fact our experiments are done on high-mobility material which shows Shubnikov-de Haas oscillations at magnetic fields as low as 100~mT and also displays clear features of the fractional quantum Hall effect it is rather counterintuitive that the density should be modulated by 50\% on the scale of microns, i.e. on a scale which is about 100 times smaller than the elastic mean free path.

Following the derivations in \cite{Efros} we can estimate how large the fluctuations of the potential can be in the studied sample. The mean of the squared random screened potential created by ionized donor atoms is expressed by the relation:
\begin{eqnarray}
\langle F^2 \rangle=W^2\frac{1}{4(q_ss)^2}~,
\label{eqn:Var}
\end{eqnarray}
where $W=\sqrt{2\pi C}e^2/\kappa$, $e$ is the electron charge, $\kappa$ is the dielectric constant, $C$ is the density of donors, $q_s=2/a_B$ is the screening parameter, $a_B=10~$nm is the Bohr radius, $s$ is the spacer width. For the studied sample, $s=70~$nm, $C\sim10^{12}~$cm$^{-2}$, $W\approx30$~meV. Thus, $\sqrt{\langle F^2 \rangle}\approx$ 1 meV. Since the spatial distribution of the charged donors, $C(r)$, is assumed to be random, not correlated, then $\langle C(r) \rangle=0$ and $\langle C(r)C(r-r') \rangle=\delta(r')$. Therefore, $\langle F \rangle^2=0$, and $\sqrt{\langle F^2 \rangle}\approx$ is nothing but the variance. The standard deviation is then $\delta F=\sqrt{\langle F^2 \rangle}\approx 1~$meV. Translating energy into carrier density, one gets: $\delta n \approx 0.3\times10^{11}~$cm$^{-2}$. From the variations of $n^*$ (\fref{fig:Spacing}) for the four studied lines one can determine that $n^*=\langle n^* \rangle \pm \delta n^* = (1.1 \pm 0.2)\times10^{11}~$cm$^{-2}$, where $\langle n^* \rangle$ is the carrier density averaged over all data points in the figure, and $\delta n^*$ is the standard deviation. The standard deviation estimated theoretically following \cite{Efros} agrees with that determined experimentally.

A similar analysis has been performed for different cooldowns of the same sample with different tips (to change the tip the sample was warmed up to room temperature and then cooled back down to base temperature). The results resemble those of \fref{fig:GateCompen} including the lobe pattern and the ring pattern. The behaviour of the fringe spacing as well as of the extracted carrier density are independent of gate voltage, thermal cycling and the tip used. They are an intrinsic property of the sample.

Interference due to direct backscattering off the tip, as assumed in the first part of scenario 1, is the most probable process to occur. Theoretical estimations of the carrier density variations agree with our experimental findings. However, due to the high purity of the studied heterostructure it is rather counterintuitive that variations of the fringe spacing are caused by inhomogeneities of the local carrier density (scenario 1) on this large scale (up to 50$\%$). The presence of sharp impurities that would lead to the angle-dependent fringe spacings is also unlikely due to the high mobility of our sample. As was also mentioned before (scenario 2), electron waves can backscatter off the tip and meet at the QPC at different angles. In this case all waves have to hit the tip under a normal angle, which is not a very probable process to occur due to the randomness of the background potential. Summing up, each scenario alone can not explain strong variations in the fringe spacing. Rather, a combination of all the physics discussed would result in a model that could account for the observed effects. In the absence of any background potential fluctuations one may expect a regular pattern of circularly shaped interference fringes. Considering the fact that several interfering paths are more prominent because of the presence of the random background potential these fringe spacings could change locally since now waves coming from different directions could interfere with each other.

When the QPC transmission is less than one, a ring pattern is expected due to the interference of waves backscattered off the tip and off the QPC (scenario~1, \fref{fig:Spacing}(d)). When the number of the QPC modes increases, more angular lobes appear, but the ring pattern remains. It is this effect that is seen in \fref{fig:GateCompen}. A similar ring pattern has been observed in the previous experiments \cite{JuraPRB1}.

We would also like to point out that the wiggles in the QPC conductance may also results from the Friedel oscillations of the electron density. The oscillations formed around the tip may reach the QPC causing the modulation of the QPC transmission, which is seen in the scanning gate images. Such a possibility is also mentioned in \cite{Szafran}. The Friedel oscillations may survive for distances on the order of a micron, which is enough to explain the appearance of the ring pattern in the gate compensated images, \fref{fig:GateCompen}. The ring pattern seems to start with a high intensity fringe followed by oscillations which decay with distance.

\section{Conclusion}

We performed scanning gate microscopy experiments of electron backscattering through a quantum point contact fabricated on a high-mobility GaAs heterostructure. A branching behaviour was observed together with interference fringes that decorate the branches. We imaged its evolution at several points on a plateau as well as between plateaus. We found that for fixed lateral position of the QPC the number and the intensity of branches change only when the QPC conductance becomes lower than an integer multiple of $2e^2/h$. When it is equal to or higher than an integer multiple of $2e^2/h$, the electron backscattering pattern remains stable. A lateral shift of the QPC, which changes the injection conditions, affects the branching pattern. The branches are thus related to the random disorder potential in the 2DEG.

Measurement of the QPC conductance as a function of the gate voltages at different tip biases, tip positions and low perpendicular magnetic fields showed that the conductance at the plateaus was restored when the B-field was applied. At the same time in the images of electron backscattering the branches of the pattern and the interference fringes gradually disappeared. This illustrated that backscattering is important for the observation of the branches and interference fringes. Some of them faded away faster than others, which means that different mechanisms of fringe formation are present.

Implementing the gating compensation technique eliminated the gating effect and enabled us to observe smaller effects due to electron backscattering. This allowed scanning closer to the QPC and increasing resolution of the SGM images. We determined the spacing between the fringes as a function of distance and found large deviations from the expected value corresponding to half of the Fermi wavelength. Several scenarios should be taken into account to explain this observation.

\section{Acknowledgements}

We are grateful for fruitful discussions with Rodolfo Jalabert, Francois Peeters, Jean-Louis Pichard, and Dietmar Weinmann. We acknowledge financial support from the Swiss National Science Foundation and NCCR ``Quantum Science and Technology".

\section*{Appendix}

\appendix
\setcounter{section}{1}

\begin{figure}[t]
\begin{center}
\includegraphics[width=15cm]{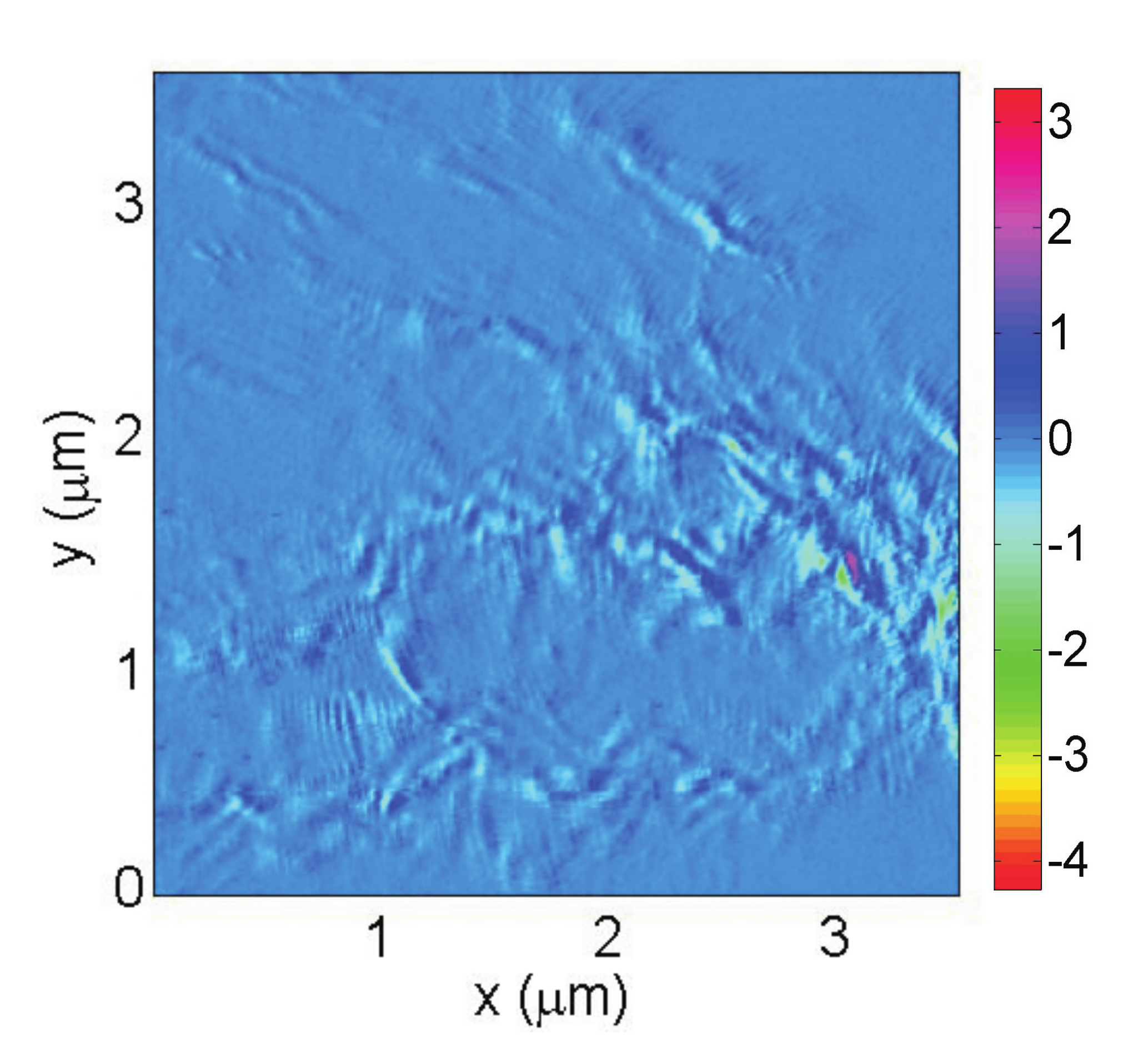}
\caption{Numerical derivative of the conductance, $\rmd G/\rmd x$, plotted in colour scale as a function of tip position. It is the same as \fref{fig:Gate}(a). Its size is increased to see the interference fringes more clearly.}
\label{fig:GateZoomin}
\end{center}
\end{figure}

\begin{figure}[t]
\begin{center}
\includegraphics[width=15cm]{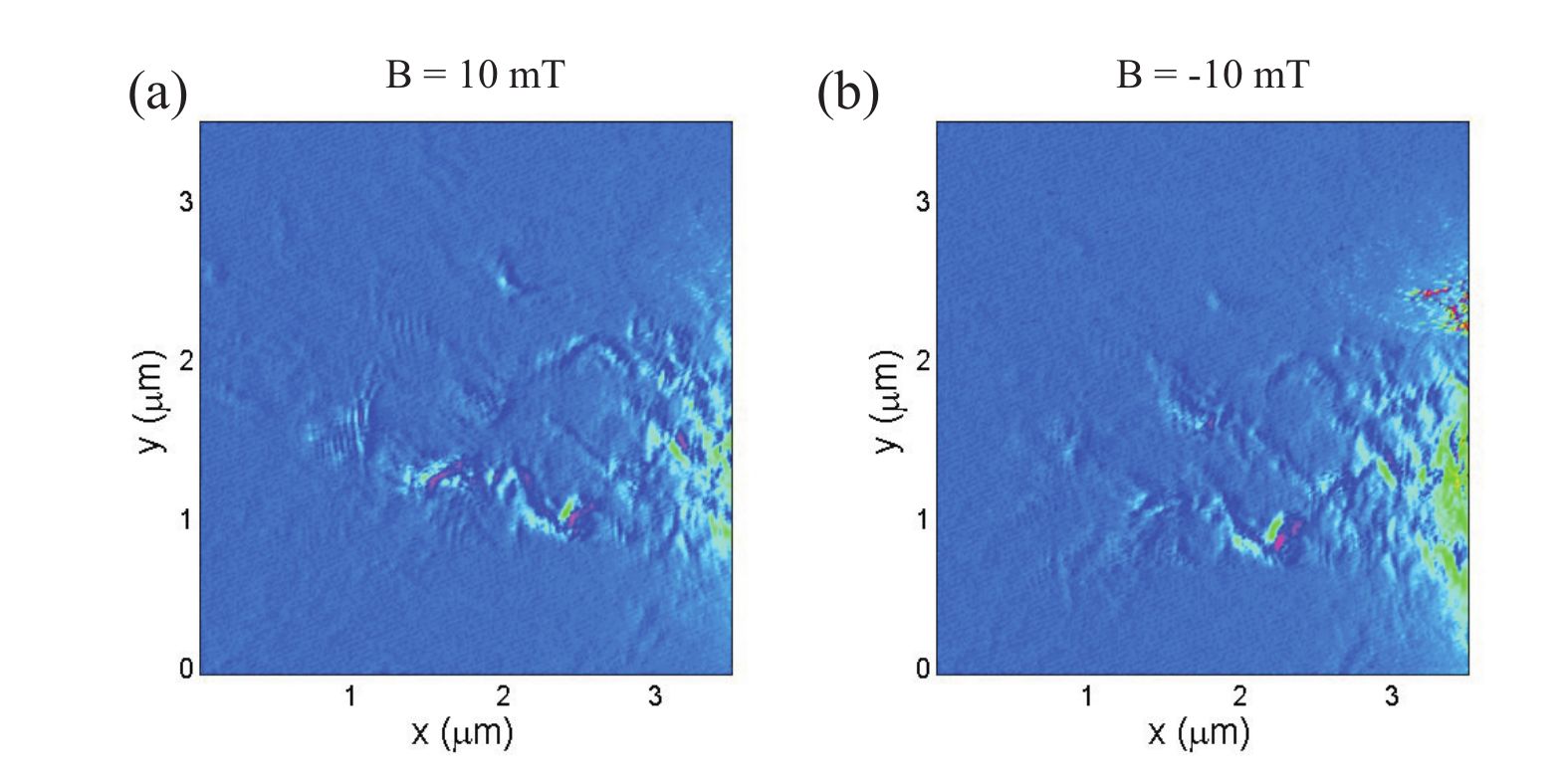}
\caption{Numerical derivative of the conductance, dG/dx, plotted in colour scale as a function of tip position for different magnetic fields. The conductance across the sample is $G=6e^2/h$. (a) B = 10 mT, (b) B = -10 mT.}
\label{fig:Bfield1}
\end{center}
\end{figure}

\begin{figure}[t]
\begin{center}
\includegraphics[width=15cm]{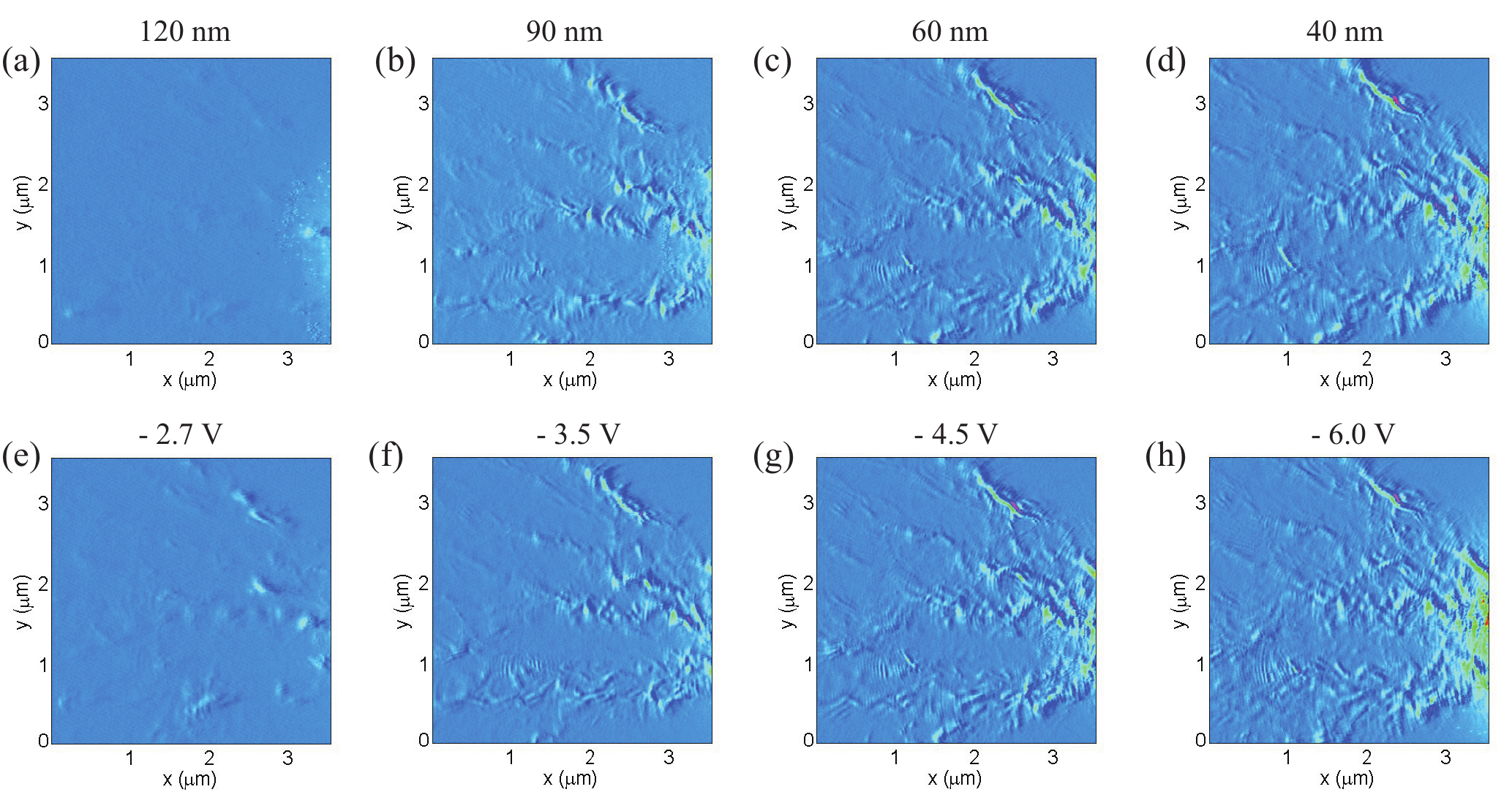}
\caption{(a)-(d) The branching behaviour as a function of the tip-surface distance: (a) 120 nm, (b) 90 nm, (c) 60 nm and (d) 40 nm. The tip bias is kept constant at -4.5 V. The QPC conductance is $6e^2/h$. (e)-(h) Electron flow as a function of the tip bias: (e) -2.7 V, (f) -3.5 V, (g) -4.5 V and (h) -6.0 V. The dip-surface distance is kept constant at 60 nm and the QPC conductance is $6e^2/h$.}
\label{fig:DistBias}
\end{center}
\end{figure}

Due to the small size of \fref{fig:Gate}, the interference fringes are not very well seen. As an example, to show that they indeed decorate the branches we increased the size of \fref{fig:Gate}(a). The result is shown in \fref{fig:GateZoomin}.

\Fref{fig:Bfield1} shows a comparison of the branching effect at opposite magnetic fields. Both images look very similar: the magnetic field suppresses electron backscattering restoring the conductance at the plateaus. Some features, which are the same in (a) and (b) seem to appear at slightly different positions. This shift can be due to the time-dependent sample drift. Branches in the upper part of the images at 10 mT are brighter than at -10 mT. The reason for this slight difference is a magnetic hysteresis: the two fields are not exactly opposite. The branches in the lower part of the images are the same, which was also observed at different positive magnetic fields in \fref{fig:Bfield} where the branches in this part of the images disappear slower than in the upper part. Thus, opposite magnetic fields affect the branching behavior in th e same way.

Electron scattering through the QPC is studied at different tip-surface distances (\fref{fig:DistBias}(a)-(d)) and tip biases (\fref{fig:DistBias}(e)-(h)). When the tip biased at -4.5V is far away above the surface, e.g. at 120 nm, it does not deplete the electron gas beneath it. There is no backscattering and therefore no branches and interference fringes. As the tip moves closer to the surface a depleted region beneath it appears. The tip-induced potential is strong enough to scatter electrons back to the constriction reducing the QPC conductance and leading to the observation of the interference fringes and the branching effect ((\fref{fig:DistBias}(b)). At smaller distances above the surface the depleted region increases and more electrons can be backscattered. This increases the width of the branches and their intensity (see \fref{fig:DistBias}(a)-(d)). A similar situation occurs when the tip bias is made more negative (see \fref{fig:DistBias}(e)-(h)).

\begin{figure}[t]
\begin{center}
\includegraphics[width=12cm]{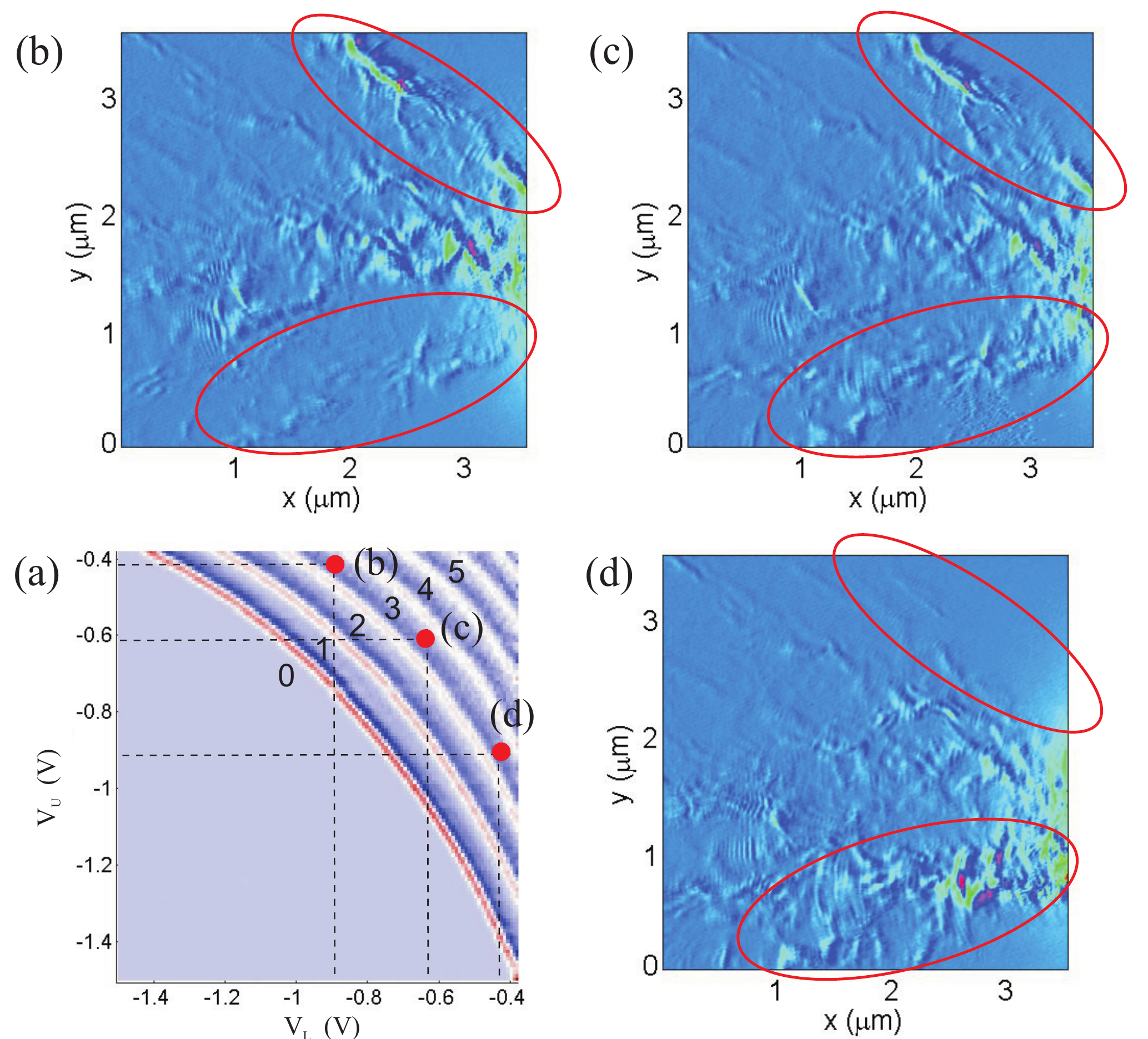}
\caption{Electron backscattering imaged as a function of unequal top gate voltages. The conductance across the sample is $G=6e^2/h$ (third plateau). (a) Transconductance, $\rmd^2G/\rmd V_\mathrm{U}\rmd V_\mathrm{L}$, as a function of the voltages, $V_\mathrm{L}$ and $V_\mathrm{U}$, applied to the lower and the upper top gates, respectively. Numbers correspond to the numbers of conductance plateaus. (b)-(d) Electron flow through the QPC at gate voltages marked by red circles in (a).}
\label{fig:Asym}
\end{center}
\end{figure}

\Fref{fig:Asym} shows the branching behaviour depending on the lateral shift of the QPC similar to what has been done in \cite{JuraNat}. In \fref{fig:Asym}(a) the transconductance $\rmd^2G/\rmd V_\mathrm{U}\rmd V_\mathrm{L}$ is plotted as a function of the voltage, $V_\mathrm{L}$, applied to the lower top gate and the voltage, $V_\mathrm{U}$, applied to the upper gate (\fref{fig:one}(b)). The light blue colour in the bottom left part of the diagram and in the rest of it corresponds to regions where the conductance through the QPC is zero or an integer multiple of $2e^2/h$, respectively. The white colour corresponds to the transition regions between the plateaus. Similar measurements were carried out on another QPC fabricated in a high mobility GaAs heterostructure \cite{Rossler} and agree with ours. Electron backscattering is studied for gate voltages marked by red circles in (a). The conductance is kept at $6e^2/h$ and a voltage applied to each of the two top gates is varied. When the QPC is shifted up (\fref{fig:Asym}(b)) by about 70 nm with respect to the symmetric case, $V_\mathrm{U}=V_\mathrm{L}$ (\fref{fig:Asym}(c)), the upper side branch (encircled) becomes more pronounced, then the lower one (encircled), and the pattern between them remains almost unchanged. In the opposite situation, when the QPC is shifted down (\fref{fig:Asym}(d)) by the same distance, the upper branch disappears and the lower one becomes wider and more intense. These images directly demonstrate the shifted injection point into the branches originating from the gate-controlled lateral shift of the QPC. No new branches appear upon the constriction shift. This means that they are fixed in space and become more/less intense and wider/narrower depending on the injection conditions, i.e. the distribution of the injected carriers. Thus, the branches are related to the disorder potential landscape in the 2DEG.

\end{document}